# Distribution of Cr atoms in a strained and strain-relaxed $Fe_{89.15}Cr_{10.75}$ alloy: Mössbauer effect study.

S. M. Dubiel[1*] and J. Żukrowski[2]

[1]AGH University of Science and Technology, Faculty of Physics and Applied Computer Science, al. A. Mickiewicza 30, 30-059 Kraków, Poland, [2]AGH University of Science and Technology, Academic Centre of Materials and Nanotechnology, al. A. Mickiewicza 30, 30-059 Kraków, Poland.

A $Fe_{89.15}Cr_{10.75}$ alloy in a heavily strained (by cold rolling) and strain-relaxed states was studied by means of the conversion electrons Mössbauer spectroscopy (CEMS). Analysis of the spectra in terms of a two-shell model revealed significant differences between the studied samples, in particular, in values of the hyperfine field, a distribution of Cr atoms within the first two neighbor-shells, the latter was expressed in terms of the short-range order parameters, and in the magnetic texture.

* Corresponding author: **Stanislaw.Dubiel@fis.agh.edu.pl**

Stainless steels constitute an important class of construction materials since a century. Their production keeps growing at the average rate of ~7 % p. a. within the last two decades. Cold working has been an essential procedure applied in the steel manufacturing industry for several reasons. First of all, cold working has been applied as a convenient strengthening method leading to an increase of the deformation resistance e. g. [1]. Next, strong plastic deformations are often involved in the industry in shaping process of metallic parts. In turn, cold work bonding is used for producing layered sheets and foils e.g. [2]. Cold worked steels are considered as appropriate materials to be used as components in the nuclear industry due to their high resistance to radiation-induced void-swelling [3]. Cold work may be also introduced unintentionally into devices of complex geometry such as vessels, tanks, pipes, heat exchangers, boilers etc. produced from sheets or plates. Cold working has, however, negative sides, too. For example, it reduces ductility and fracture toughness [4] as well as the fatigue life of components manufactured from a pre strained materials [3]. In terms of industrial production and macroscopic properties of stainless steels it is of vital importance to establish relationship on one side between microstructure and texture in metal working processes and the process conditions such as strain, strain rate, temperature of deformation and annealing on the other side. It is well known that during cold working, in addition to dislocations, the following lattice defects are produced: stacking faults, stacking faults bundles, deformation twins, deformation bands. Precipitates of $\alpha'$ and $\varepsilon$ martensites may be also formed [5]. The recovery of the cold-rolled products is mainly attributed to the annihilation and rearrangements of dislocations [6]. However, the lattice is predominantly composed of atoms not from defects. In the stainless steels the second most abundant (and important) are Cr atoms. Do these Cr atoms experience any kind of rearrangements during the recovery of the strain induced by the cold work? This is a question we wanted to answer by studying a $Fe_{89.15}Cr_{10.75}$ model alloy in two different metallurgical states viz. a heavily strained by cold rolling and a strain-relaxed samples.

The chemical composition of the studied alloy, as revealed by the microprobe analysis, is displayed in Table 1.

Table 1

Chemical composition of the investigated alloy.

| Cr [wt. pct.] | C [wt. ppm] | S [wt. ppm] | O [wt. ppm] | N [wt. ppm] | P [wt. ppm] |
|---|---|---|---|---|---|
| 10.10 | 4 | 6 | 4 | 3 | < 5 |

The studied sample was in form of a ~30 μm thick ribbon obtained from the original rod by cold rolling a ~1 mm thick disc cut off the rod. For the study six ~20 x ~20 mm² squares were cut from the ribbon. Three samples, called hereafter, strained, were studied without any further treatment, while the remaining three ones were vacuum (~$10^{-5}$Torr) annealed at 700°C for 1 h followed by a furnace cooling down to room temperature (RT). These samples are referred to hereafter as strain-free.

The samples were measured at RT by means of the Mössbauer spectroscopy. Recorded were conversion electrons (CEMS mode) on each side of the samples, marked as A and B, in Figs.2 and 3 . The CEMS spectra contain information on ~0.3µm thick pre surface zone. An example of the spectra measured on the strained and the strain-free samples is shown in Fig. 1. Optically, a difference between the two spectra can be hardly seen. However, a distribution curves of the hyperfine field derived from the two spectra exhibit pronounced difference.

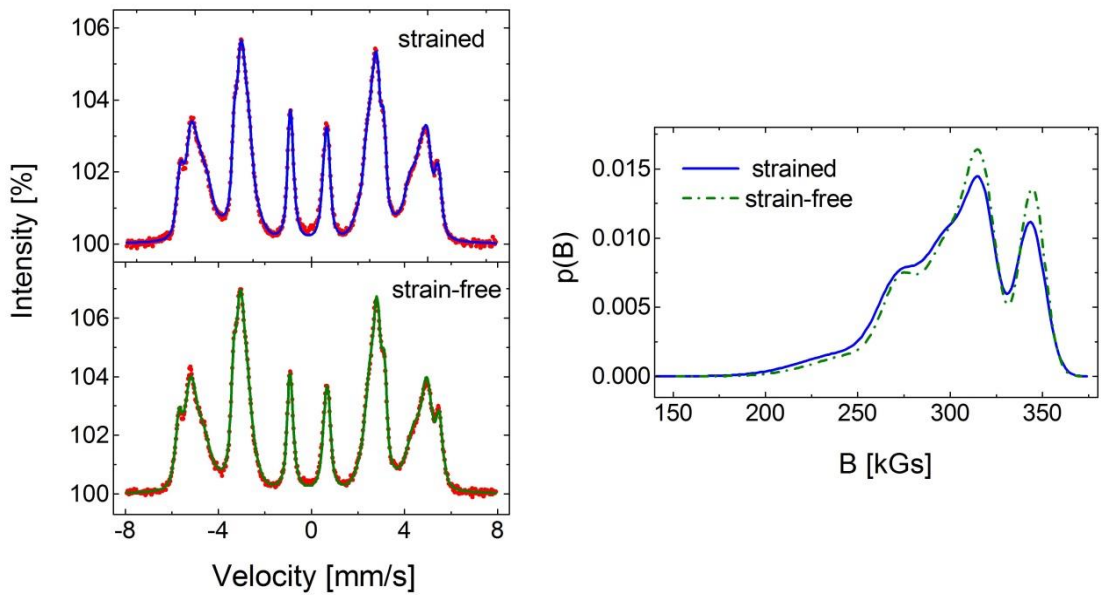

Fig. 1 $^{57}$Fe CEMS spectra recorded at RT on strained and strain-free samples of the $Fe_{89.15}Cr_{10.75}$ alloy. The right-hand panel shows the distributions of the hyperfine field derived from the spectra.

The measured spectra were analyzed using two procedures: (I) Hyperfine magnetic field distribution method as outlined elsewhere [7], and (II) Two-shell superposition method outlined in detail elsewhere [8,9]. Concerning (I), a linear relationship between the hyperfine field and the isomer shift was assumed following the experimental finding [8]. The fitting procedure (I) yielded hyperfine magnetic field distribution curves, examples of which can be seen in Fig.1 (right-hand panel). Their integration gave an average value of the hyperfine field, $\langle B \rangle = \int p(B)dB$. The procedure (B) enabled a quantitative insight into a distribution of Cr atoms within the first-two neighbor shells around the probe Fe atoms. This chance follows from the fact that the Fe-site hyperfine field, $B$, is not only sensitive to Cr atoms present in the first (1NN) and in the second (2NN) shells but even a clear cut distinction between Cr atoms situated in *1NN* and those in *2NN* can be made [8,9]. Thanks to this property, the analysis of the Mössbauer spectra in terms of method (II) gave an information on probabilities of various atomic configurations, *P(m,n)*, *m* being a number of Cr atoms in *1NN*, and *n* that in *2NN*. Knowing the atomic configurations, *(m,n)* and the *P(m,n)*-values enables, in turn, calculation of an average number of Cr atoms in *1NN*, *<m>=Σm·P(m,n)*, that in *2NN*, *<n>=Σn·P(m,n)*, as well as in both shells, *<m+n> = <m>+<n>*. In turn, the knowledge of *<m>*, *<n>* and *<m+n>* found in that way combined with the corresponding quantities expected for the random distribution, *<m>$_r$=8x*, *<n>$_r$=6x*, and *<m+n>$_r$=14x*, permits determination of short-range order (SRO) parameters. For this purpose the following formulae can be used:

$$\alpha_1 = 1 - \frac{\langle m \rangle}{\langle m_r \rangle} \qquad (1a)$$

$$\alpha_2 = 1 - \frac{\langle n \rangle}{\langle n_r \rangle} \qquad (1b)$$

$$\alpha_{12} = 1 - \frac{\langle m + n \rangle}{\langle m_r + n_r \rangle} \qquad (1c)$$

In practice, the spectra were analyzed assuming the effect of presence of Cr atoms in the *1NN-2NN* vicinity of the $^{57}$Fe probe nuclei on the hyperfine field, *B*, and the isomer shift, *IS*, was additive i.e. $X(m,n) = X(0,0) + m\Delta X_1 + n\Delta X_2$, where *X=B* or *IS*, *ΔX$_k$* is a change of *X* due to one Cr atom situated in *1NN* (*k*=1) and in *2NN* (*k*=2). The total number of possible atomic configurations *(m,n)* within the 1NN-2NN approximation is equal to 63, but for *x* = 10.75 at% most of them have vanishingly small probabilities, so only 10 the most probable (according to the binomial distribution) were included into the fitting procedure (their overall probability was > 0.99). However, their probabilities (related to spectral areas of sextets corresponding to the chosen configurations) were treated as free parameters. Free parameters were also *X(0,0)*, line widths and their relative ratios (C1:C2:C3). On the other hand, fixed were values of $\Delta X_k$'s viz. $\Delta B_1$= -30.5 kOe, $\Delta B_2$= -19.5 kOe, $\Delta IS_1$ = -0.02 mm/s, and $\Delta IS_1$= -0.01 mm/s [8,9,10].

The best-fit spectral parameters and calculated therefrom average ones are displayed in Table. 2.

**Table 2**

The best-fit spectral parameters as derived from CEMS spectra recorded on both sides (A and B) of the investigated samples of a FeCr$_{10.75}$ alloy. The meaning of the symbols is given in the text. P(0,0) is in percentage, Isomer shifts are in mm/s (relative to the Co/Rh source at RT) and hyperfine fields in kOe.

| Sample | P(0,0) | IS(0,0) | \<IS\> | B(0,0) | \<B\> | \<m\> | \<n\> | C2/C3 | θ [°] |
|---|---|---|---|---|---|---|---|---|---|
| **Strained** | | | | | | | | | |
| 1 A | 19.7 | -0.105 | -0.127 | 343.8 | 302.1 | 0.897 | 0.734 | 3.03 | 72.5 |
| 1 B | 20.3 | -0.103 | -0.128 | 344.3 | 302.9 | 0.902 | 0.711 | 3.10 | 73.8 |
| 2 A | 19.5 | -0.103 | -0.128 | 342.9 | 301.0 | 0.889 | 0.759 | 2.97 | 71.9 |
| 2 B | 20.8 | -0.104 | -0.130 | 345.7 | 304.1 | 0.922 | 0.691 | 2.95 | 71.5 |
| 3 A | 19.5 | -0.103 | -0.128 | 344.3 | 302.4 | 0.885 | 0.763 | 3.01 | 72.3 |
| 3 B | 19.1 | -0.102 | -0.127 | 344.2 | 305.8 | 0.824 | 0.696 | 3.05 | 73.4 |
| **Strain-free** | | | | | | | | | |
| 1 A | 22.2 | -0.103 | -0.127 | 344.5 | 305.8 | 0.824 | 0.696 | 3.24 | 75.3 |
| 1 B | 22.9 | -0.104 | -0127 | 345.1 | 307.3 | 0.808 | 0.678 | 3.30 | 76.0 |
| 2 A | 21.3 | -0.104 | -0.127 | 345.6 | 306.7 | 0.823 | 0.707 | 3.25 | 75.4 |
| 2 B | 22.7 | -0.105 | -0.129 | 346.0 | 307.8 | 0.811 | 0.690 | 3.25 | 75.4 |
| 3 A | 22.3 | -0.105 | -0.128 | 345.1 | 307.0 | 0.798 | 0.704 | 3.29 | 76.0 |
| 3 B | 21.9 | -0.103 | -0.126 | 343.9 | 305.4 | 0.798 | 0.722 | 3.26 | 75.5 |

Already a visual inspection of Table 2 gives evidence on meaningful differences between the strained and strain-free samples. Let us first consider average values of some quantities that differ significantly. Firstly, the average value of *P(0,0)* i.e. the probability of an atomic configuration with no Cr atoms within the 1NN-2NN neighborhood is equal to 19.8% for the strained sample and to 22.2% for the strain-free. Alike difference exists in the average values of the hyperfine field, namely <B>=303.1 kOe and 306.7 kOe, respectively. The higher the <*B*>-value the lower the Cr concentration and the number of Cr atoms in the vicinity of the probe Fe probe atoms [8,9]. The difference of 3.6 kOe corresponds to 1.3 at. % Cr. The different distribution of Cr atoms in the two types of the samples is also reflected in the average numbers of Cr atoms in 1NN, <m>, and in 2NN, <n>, neighbor-shells. For the strained samples <m>=0.8865 and <n>=0.7260, whereas for the strain-free samples the corresponding figures are 0.8100 and 0.6995, respectively. For the case of random distribution <m>$_r$=0.860 and <n>$_r$=0.6450. It is obvious that the actual distribution is not random. The departure from randomness can be also expressed in terms of a local concentration of Cr, $x_k(at.\%) = \frac{<n_k>}{M} 100$, where *M*=8, 6, 14 for *k*=1 (1NN), 2 (2NN), and 12 (1NN+2NN), respectively.  In this way one gets $x_1$=11.1, $x_2$=12.1 and $x_{12}$=11.5 at.% for the strained sample, and $x_1$=10.1, $x_2$=11.65 and $x_{12}$=10.8 at.% for the strain-free one. The effect of strain and its relaxation was also revealed in the magnetic texture measured in this experiment by an average angle, θ, between the magnetization vector in the pre surface zone and the normal to the surface. The data displayed in Table 2 indicate a 3$^o$ difference in θ viz. the strain-relaxation treatment obviously caused a 3$^o$ rotation of the magnetization vector towards the surface.

The above-discussed difference in the distribution of Cr atoms can be followed for each sample and neighbour-shell individually. For this purpose values of the SRO-parameters were calculated based on equations (1a)-(1c). They are presented in Fig. 2.

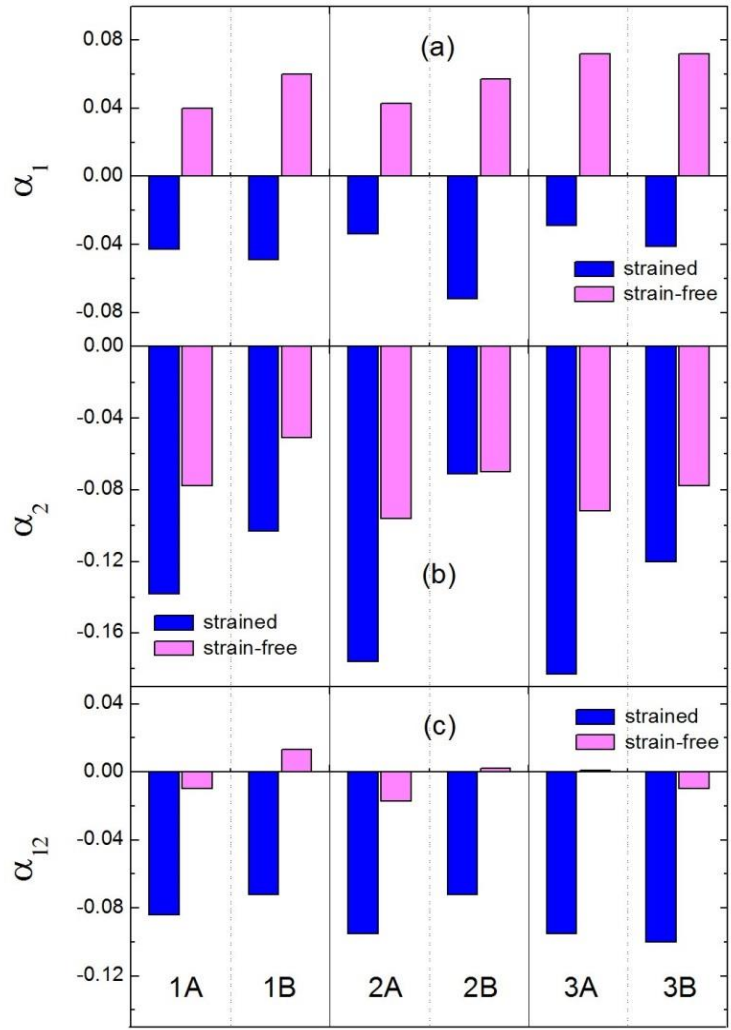

Fig. 2 SRO parameters for the three samples of the Fe-Cr alloy.

Here the contrast between the strained and strain-relaxed samples is clearly visible. Interestingly, the behavior is different for 1NN and 2NN shells viz. the number of Cr atoms in 1NN is for the strained sample higher and for the strain-free sample it is lower than the one expected for the random case. The 2NN shell is overpopulated by Cr atoms, yet the overpopulation in the strained sample is significantly higher. The population of Cr atoms averaged over both shells is practically random for the strain-free sample and highly overpopulated in the strained sample.

Interesting and noteworthy feature revealed in this study and shown in Fig. 2 is heterogeneity in the distribution of Cr atoms seen not only for individual neighbor shells but also for different samples cut out of the same piece of the rolled ribbon, and even for different sides of a given sample. The differences exist even at the level of individual atomic configurations, and, in particular, in the most probable one i.e. *(0,0)*. Its probabilities, *P(0,0)*, can be converted into a local atomic concentration, *x(0,0)*, using a binomial approximation viz. $x(0,0) = 1 - \sqrt[14]{P(0,0)}$. The output of such procedure is presented in Fig. 3.

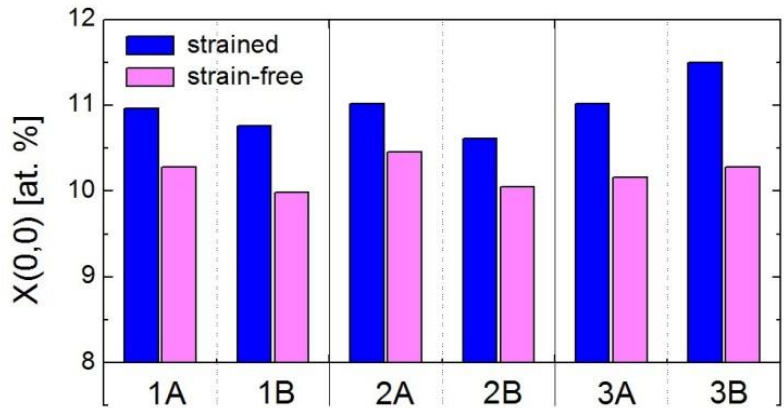

Fig. 3 Local concentration of Cr, *x(0,0)*, derived from the probability of the *P(0,0)* atomic configuration in the binomial approximation.

The average value of *x(0,0)* for the strained sample is 11.0 at. % i.e. slightly more than average concentration obtained by the chemical analysis, and 10.2 at% for the strained-free sample. However, the strained sample seems to be more heterogeneous as a fluctuation in *x(0,0)* reaches ~1 at.% while the one in the strain-free amounts to ~0.5 at.%.

In summary, significant differences in the distribution of Cr atoms within the first (1NN) and the second (2NN) neighbor shells were revealed for the strained and the strain-relaxed samples of a $Fe_{89.15}Cr_{10.75}$ alloy. This finding evidently shows that a rearrangement of Cr atoms has also to be taken into account in order to properly understand the recovery of the cold worked stainless steels. The results reported in this paper seem to be of a high importance as a better understanding of the deformation of these steels has great practical importance with regard both to the formability of these materials as well as to the optimization of their properties.

**Acknowledgements**

This work has been carried out within the framework of the EUROfusion Consortium and has received funding from the European Union's Horizon 2020 research and innovation program under grant agreement number 633053. The views and opinions expressed herein do not necessarily reflect those of the European Commission. This work was supported by the Ministry of Science and Higher Education (MNiSW), Warsaw and the AGH University of Science and Technology, Krakow, Poland.